\documentclass[aps,pre,twocolumn,superscriptaddress,floatfix]{revtex4}

\usepackage{amsmath,amssymb,amsfonts}
\usepackage{graphicx}
\usepackage{xeCJK}
\usepackage[colorlinks=true,linkcolor=blue,citecolor=blue,urlcolor=blue]{hyperref}
\usepackage{bm,braket}

\begin{document}

\title{Von Neumann's Two Quantum Entropies}
\author{Biao Wu(吴飙)}
\email{wubiao@pku.edu.cn}
\affiliation{International Center for Quantum Materials, School of Physics, Peking University, Beijing 100871, China}
\affiliation{Wilczek Quantum Center, Shanghai Institute for Advanced Studies, University of Science and Technology of China, Shanghai 201315, China}
\affiliation{Hefei National Laboratory, Hefei 230088, China}
\affiliation{Beijing Key Laboratory of Quantum Devices, Peking University, Beijing 100871, China}

\date{\today}

\begin{abstract}
Von Neumann proposed two quantum entropies, one in 1927 and the other in 1929.
He intended to define thermodynamic entropy for quantum states in both cases.
The 1927 entropy is the well-known von Neumann entropy, which is zero for all pure states.
The 1929 entropy, which is nonzero for almost all pure states, has largely been ignored.
After reviewing the history and properties of both entropies, we argue that the 1927 entropy is 
not thermodynamic entropy and should be used simply as a measure of entanglement, 
whereas the 1929 entropy is the true thermodynamic entropy for quantum states.
\end{abstract}

\maketitle

\section{Introduction}

Thermodynamic entropy was developed for macroscopic systems to explain heat-related irreversible phenomena.
Since macroscopic systems are made of atoms or molecules, it is natural to seek a microscopic understanding of thermodynamic entropy.
The first major step was made by Boltzmann, who derived the $H$ theorem.
Although the derivation required an assumption, it can partially explain the origin of entropy and why it always increases in an isolated system~\cite{Huang}.

In 1926, it became clear that microscopic particles follow quantum dynamics, and one needs to understand entropy in terms of quantum mechanics.
Von Neumann made his first attempt in 1927~\cite{Neumann1927,Duncan}.
For a quantum system with density matrix $\hat{\rho}$, he found that its thermodynamic entropy is
\begin{equation}\label{eq:vNentropy}
S_{v} = -\operatorname{Tr}(\hat{\rho} \ln \hat{\rho})\,,
\end{equation}
which is the well-known von Neumann (vN) entropy. In this work we always set $k_B=1$.
In 1929~\cite{Neumann1929}, he defined another entropy for quantum systems,
\begin{equation}\label{eq:WvNentropy}
S_{w} = -\sum_{\alpha} p_{\alpha} \ln p_{\alpha}\,,
\end{equation}
where $p_{\alpha}$ is a coarse-grained probability distribution over a phase space divided into Planck cells.
As von Neumann stated in his paper, this definition of entropy was originally due to Wigner; 
we therefore call it the Wigner--von Neumann (WvN) entropy to distinguish it from the well-known von Neumann entropy.
Von Neumann was able to prove a quantum $H$ theorem with the WvN entropy.
One crucial difference is that the vN entropy is always zero for a pure quantum state, whereas the WvN entropy is nonzero for almost all pure quantum states.
The English translation of the 1929 paper can be found in Ref.~\cite{von2010proof}.

We briefly review the history and derivations of these two quantum entropies and examine their properties.
In the end, we argue that the vN entropy is a measure of entanglement, while the WvN entropy is the thermodynamic entropy for quantum systems.

\section{Von Neumann Entropy}\label{sec:vN}
\subsection{Von Neumann's derivation}
Von Neumann published three papers in 1927~\cite{Duncan}, which examined the mathematical 
foundations of quantum mechanics.
The von Neumann entropy was defined in the third paper.
These three papers were expanded into the well-known book \textit{Mathematical Foundations of Quantum Mechanics}~\cite{vonNeumannBook}.
We briefly review the derivation according to this book.

\begin{figure}[htbp]
\centering
\includegraphics[width=0.45\textwidth]{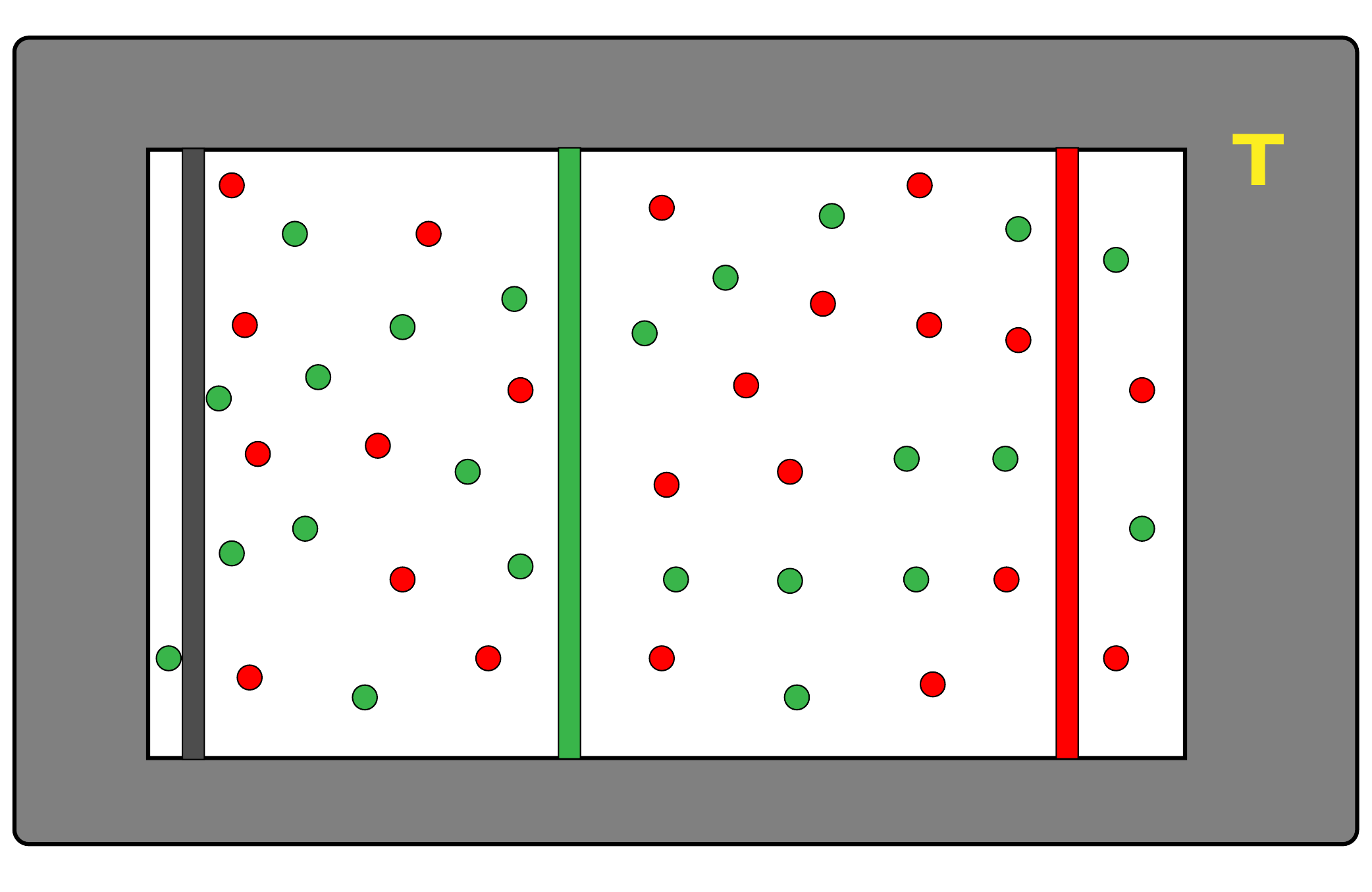}
\caption{
Von Neumann's ideal gas of fictitious molecules. 
    For a simple density matrix in Eq.~(\ref{eq:rho_mix}), the gas consists of two species: green molecules 
    in state $\ket{\psi_1}$ and red molecules in state $\ket{\psi_2}$. ​
To separate them, von Neumann imagined three kinds of walls. A black wall is impenetrable 
to both species. A green wall acts as a semi-permeable membrane: it lets green molecules pass 
but reflects red ones. A red wall does the opposite, admitting red molecules 
while reflecting green ones. The entire box sits in contact with a thermal bath at temperature $T$.
}
\label{fig:gas}
\end{figure}

Von Neumann first argued that all pure states should have the same entropy, which is zero.
To find the entropy of a mixed state, von Neumann designed a thought experiment with an ideal gas consisting of fictitious molecules, each of which is a copy of the quantum system.
For simplicity, we consider the density matrix
\begin{equation}\label{eq:rho_mix}
\hat{\rho} = \frac{1}{2}\bigl(|\psi_{1}\rangle\langle\psi_{1}| + |\psi_{2}\rangle\langle\psi_{2}|\bigr)\,,
\end{equation}
where $|\psi_{1}\rangle$ and $|\psi_{2}\rangle$ are two orthogonal quantum states of a given system.
In this case, von Neumann's ideal gas has two kinds of fictitious molecules, 
one in state $|\psi_{1}\rangle$ and the other in $|\psi_{2}\rangle$, as shown in Fig.~\ref{fig:gas}.

The key to von Neumann's argument is that one can construct semi-permeable walls, e.g., the green wall in Fig.~\ref{fig:gas}.
For this wall, a green molecule representing state $|\psi_{1}\rangle$ can pass through, while red molecules are bounced off.
Von Neumann argued that there exists in principle an operator $\hat{R}$ such that
\begin{equation}
\hat{R}|\psi_{j}\rangle = a_{j}|\psi_{j}\rangle\,, \qquad j = 1,2\,.
\end{equation}
As a result, one can place instruments in the wall that make measurements of $\hat{R}$.
When a molecule hits the wall, the instruments capture the molecule and make a measurement.
For the green wall, when the result is $a_{1}$, the instrument lets the molecule pass through 
with its hitting velocity; otherwise, the instrument rejects the molecule as if the molecule had undergone an elastic bounce.

\begin{figure}[htbp]
\centering
\includegraphics[width=0.45\textwidth]{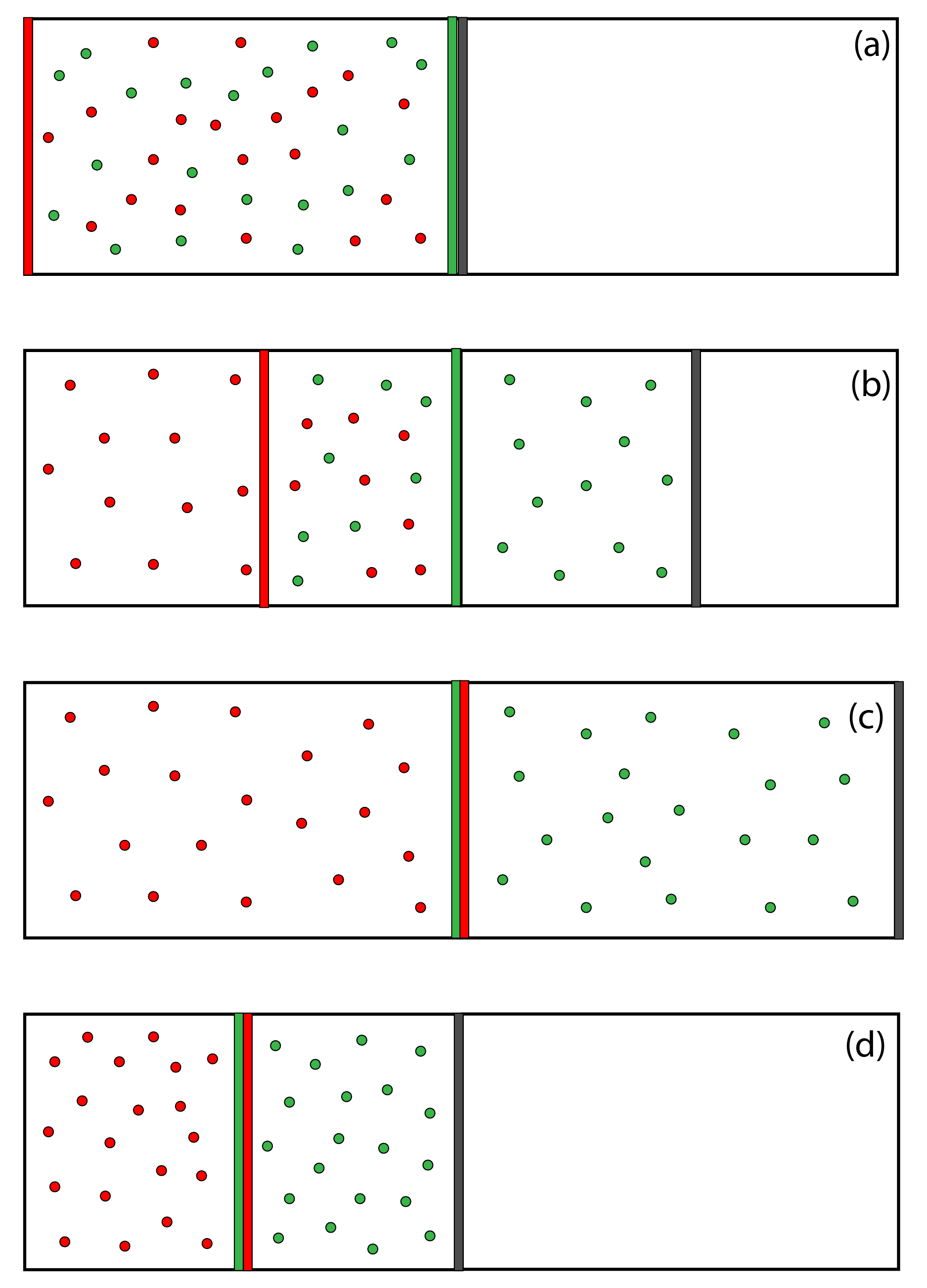}
\caption{Separating the molecules in von Neumann's ideal gas. (a) An empty box identical to 
the original is placed to its right. A green semi-permeable wall and a black impenetrable wall 
are inserted between the two boxes, while a red semi-permeable wall is fixed at the left end 
of the original box. (b) The black and red walls move slowly to the right, 
maintaining their separation. (c) When the black wall reaches the right end 
of the original box, the gas is fully separated: red molecules occupy the left box 
and green molecules the right. Because the walls move quasi-statically, 
no entropy is generated. (d) In the final stage, all three walls shift leftward in unison. 
The two semi-permeable walls halt at the center of the left box, 
while the black wall continues moving until it comes to rest between the two boxes. 
The right-hand box is now empty and can be removed. For clarity, 
the thermal bath at temperature $T$  is not shown.}
\label{fig:sep}
\end{figure}

Using two semi-permeable walls and one completely impenetrable wall, one can reversibly separate the molecules.
This is illustrated in Fig.~\ref{fig:sep}.
The procedure has three steps:
(1)~An identical empty box is placed to the right of the original box.
The two boxes are separated by an impenetrable wall and a green semi-permeable wall, while a red semi-permeable wall is placed at the left end of the original box.
(2)~The red semi-permeable wall and the impenetrable wall move slowly to the right while keeping their distance fixed.
When the black wall arrives at the left end of the right box, the molecules are separated, with the red ones in the left box and the green ones in the right box.
(3)~Finally, the three walls move simultaneously and slowly to the left while keeping the distances between them fixed.
As a result, the volume of the red molecules is compressed.
When the compression reaches 50\%, the two semi-permeable walls stop, but the black wall continues to move to the left to reduce the volume of the green molecules.
When the compression reaches 50\%, the black wall stops.
At the end, the right box becomes empty again and can be removed.

During the whole procedure, the entropy of the ideal gas changes only during the final stage, when the volumes of red and green molecules are compressed.
Since the boxes are kept in contact with a heat bath at temperature $T$, the work done by the walls is transformed into heat that is transferred to the heat bath.
The change of entropy equals the work divided by $T$, which is
\begin{equation}
\Delta S = N_{1} \ln 2 + N_{2} \ln 2\,.
\end{equation}
As each molecule is a copy of the quantum system, the entropy for the quantum system is
\begin{equation}
S = \frac{\Delta S}{N} = -\frac{1}{2}\ln\frac{1}{2} - \frac{1}{2}\ln\frac{1}{2}\,,
\end{equation}
which is just the entropy in Eq.~(\ref{eq:vNentropy}) for the special density matrix~(\ref{eq:rho_mix}).

The above derivation is in a sense mysterious because it shows that the thermodynamic entropy of a quantum system 
is equal to the mixing entropy of a classical ideal gas of fictitious molecules.
It is not clear at all why this has anything to do with the actual heat and disorder in a real quantum system.
Moreover, the development of quantum computing has made physicists aware that constructing 
semi-permeable walls capable of measuring the operator $\hat{R}$ is physically unfeasible.
Mathematically, it is easy to formulate $\hat{R}$, for example, 
as $\hat{R}=a_{1}|\psi_{1}\rangle\langle\psi_{1}|+a_{2}|\psi_{2}\rangle\langle\psi_{2}|+\cdots$.
Physically, constructing such an instrument to measure $\hat{R}$ 
in general requires an exponentially large amount of physical resources (see discussion in Sec.\ref{sec:basis}).

\subsection{Everett's Insight}
In his 1957 PhD thesis~\cite{Everett}, on his way to proposing the many-worlds theory, Everett found that 
the von Neumann entropy could be used to quantify the entanglement between two subsystems.
Everett considered a composite quantum system consisting of two subsystems with Hilbert spaces $\mathcal{H}_{1}$ and $\mathcal{H}_{2}$.
Let $\hat{A}$ with eigenstates $|\psi_{n}\rangle$ be a Hermitian operator on $\mathcal{H}_{1}$, 
and $\hat{B}$ with eigenstates $|\phi_{n}\rangle$ be a Hermitian operator on $\mathcal{H}_{2}$.
A given quantum state $|\Psi\rangle$ of the composite system can be expanded as
\begin{equation}
|\Psi\rangle = \sum_{m,n} a_{mn} |\psi_{m}\rangle \otimes |\phi_{n}\rangle\,.
\end{equation}
For this state, Everett defined the correlation between the two subsystems as
\begin{equation}
C_{\hat{A}\hat{B}} = \sum_{m,n} p_{mn} \ln p_{mn} - \sum_{m} p_{m} \ln p_{m} - \sum_{n} p_{n} \ln p_{n}\,,
\end{equation}
where $p_{mn}=|a_{mn}|^{2}$, $p_{m}=\sum_{n}p_{mn}$, and $p_{n}=\sum_{m}p_{mn}$.
Everett conjectured that this correlation $C_{\hat{A}\hat{B}}$ becomes maximal 
when $\hat{A}$ is the reduced density matrix $\hat{\rho}_{A}=\operatorname{Tr}_{B}(|\Psi\rangle\langle\Psi|)$ 
and $\hat{B}$ is $\hat{\rho}_{B}=\operatorname{Tr}_{A}(|\Psi\rangle\langle\Psi|)$.
Everett called it canonical correlation, which is exactly the von Neumann entropy, 
$-\operatorname{Tr}(\hat{\rho}_{A}\ln\hat{\rho}_{A})=-\operatorname{Tr}(\hat{\rho}_{B}\ln\hat{\rho}_{B})$.
The conjecture was later proved by Donald.
Everett's PhD thesis is long and loaded with notions unfamiliar to modern readers; 
a bridged version with modern notation can be found in Ref.~\cite{Wu2021MW}.

This canonical correlation, as called by Everett, is precisely entanglement.
Everett's result was eventually noticed by Schumacher~\cite{Schumacher1995}, 
and the von Neumann entropy began to be used to measure entanglement.
Unfortunately, Everett was cited only once by Schumacher in 1995~\cite{Schumacher1995}, 
and his contribution was later completely ignored by the quantum-information community~\cite{H4rmp}.

\section{Wigner--von Neumann Entropy}\label{sec:WvN}

In 1929, in an attempt to prove a quantum $H$ theorem, von Neumann found the 1927 
entropy inapplicable and had to introduce a different entropy.
Here is the reason in von Neumann's own words:
``The expressions for entropy given by the author in [1927] are not applicable here in the way they were intended, 
as they were computed from the perspective of an observer who can carry out all measurements that are possible in principle \dots\ 
If we take into account that the observer 
can measure only macroscopically then we find different entropy values.''~\cite{von2010proof}

\begin{figure}[htbp]
\centering
\includegraphics[width=0.35\textwidth]{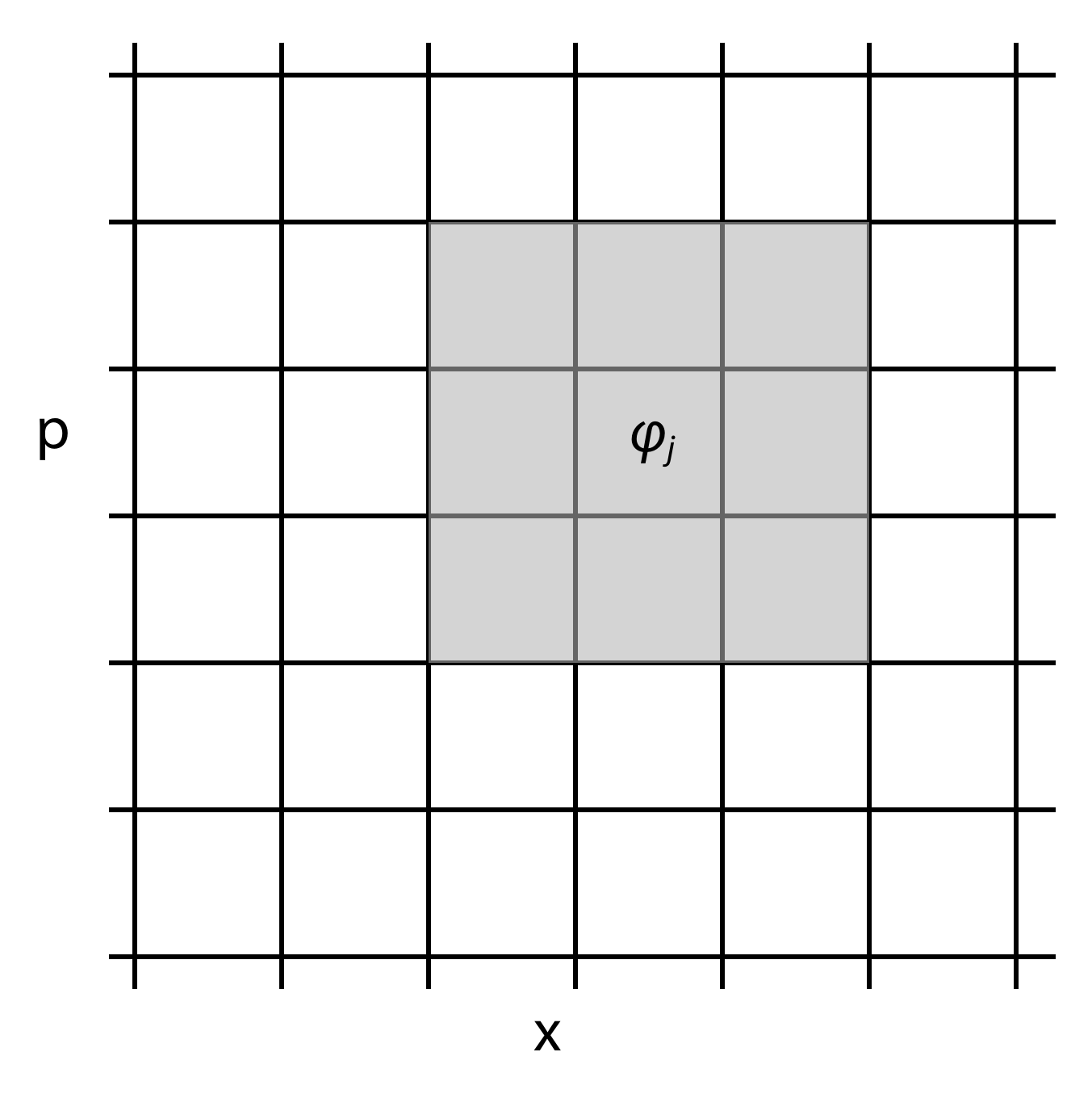}
\caption{Phase space partitioned into Planck cells. Each cell is represented 
    by a localized wave function $\ket{\varphi_i}$​. The shaded region depicts 
    a single macroscopic state, which encompasses many Planck cells---each 
    one a distinct microscopic state.}
\label{fig:phase}
\end{figure}

For a classical phase space, it can be divided into Planck cells as shown in Fig.~\ref{fig:phase}.
Von Neumann argued that one could assign a localized wave function $|\varphi_{j}\rangle$ to each 
Planck cell and that these wave functions form a complete orthonormal basis.
As a result, for a given quantum state $|\psi\rangle$ of the system, we have the expansion
\begin{equation}
|\psi\rangle = \sum_{j} |\varphi_{j}\rangle \langle\varphi_{j}|\psi\rangle\,,
\end{equation}
which gives a probability distribution $p_{j}=|\langle\varphi_{j}|\psi\rangle|^{2}$ over phase space.
After coarse graining, as in Fig.~\ref{fig:phase}, one obtains a new probability distribution $p_{\alpha}$ 
and the new entropy in Eq.~(\ref{eq:WvNentropy}).
Von Neumann claimed that the idea of this new entropy came from Wigner~\cite{von2010proof}, 
and so we call it the Wigner--von Neumann (WvN) entropy.

However, the WvN entropy has been largely forgotten along with the quantum $H$ theorem.
There has been discussion as to why von Neumann's 1929 paper was forgotten~\cite{Goldstein2010}.
In our opinion, there are at least two technical reasons that the WvN entropy 
has not been widely adopted:
(1)~It was not clear how many microscopic states correspond to one macroscopic state, i.e., 
how the coarse graining could be carried out in a practical computation.
(2)~The method suggested by von Neumann to compute $|\varphi_{j}\rangle$ is time-consuming 
and ``cumbersome'' in von Neumann's own words~\cite{von2010proof}.
These two technical issues have been resolved.
An efficient numerical method has been developed to find $|\varphi_{j}\rangle$ in the form of Wannier functions $|w_{j}\rangle$~\cite{Fang2018,Huang2026}.
Furthermore, coarse graining is found to be unnecessary~\cite{Han2015}: with the entropy defined as
\begin{equation}\label{eq:WvN_noCG}
S_{w} = -\sum_{j} |\langle w_{j}|\psi\rangle|^{2} \ln |\langle w_{j}|\psi\rangle|^{2}\,,
\end{equation}
one can still prove the quantum $H$ theorem~\cite{Han2015}.

Both the vN entropy and the WvN entropy were introduced by von Neumann as thermodynamic entropy.
Their differences are quite obvious.
The vN entropy is always zero for pure quantum states and does not change under unitary evolution, whereas 
the WvN entropy is nonzero for almost all pure quantum states and changes under unitary evolution.
In his famous book \textit{Mathematical Foundations of Quantum Mechanics}, von Neumann 
had a lengthy discussion about these differences~\cite{vonNeumannBook}.
Von Neumann first noted that it is ``surprising'' that the vN entropy is invariant under unitary time evolution.
To explain this surprise, he made an analogy to a classical gas.
For a classical gas, if one knows the position and momentum of every particle, the entropy of the gas is zero because one has full knowledge of the gas.
Therefore, for a quantum pure state, the observer who ``can find out (measure) everything which is measurable in principle'' knows everything about the system.
As a result, the vN entropy of the system is zero and stays at zero even when the system evolves to a different pure state.
To have an entropy that changes with time, as observed in experiments, one should have an entropy 
``as seen by an observer who cannot measure all quantities, but only a few special quantities, namely, the macroscopic ones.''

Therefore, according to von Neumann, for the thermodynamic entropy that is observed and measured in macroscopic systems, we should use the WvN entropy.
Instead, and interestingly, people have been using the vN entropy and have not found any inconsistency.
We will explain that this is due to a mathematical coincidence.

\section{Hindsight}\label{sec:hindsight}

Looking back after almost 100 years, we think that the vN entropy should be used simply 
as a measure of quantum entanglement and the WvN entropy as the thermodynamic entropy.

\begin{figure}[htbp]
\centering
\includegraphics[width=0.4\textwidth]{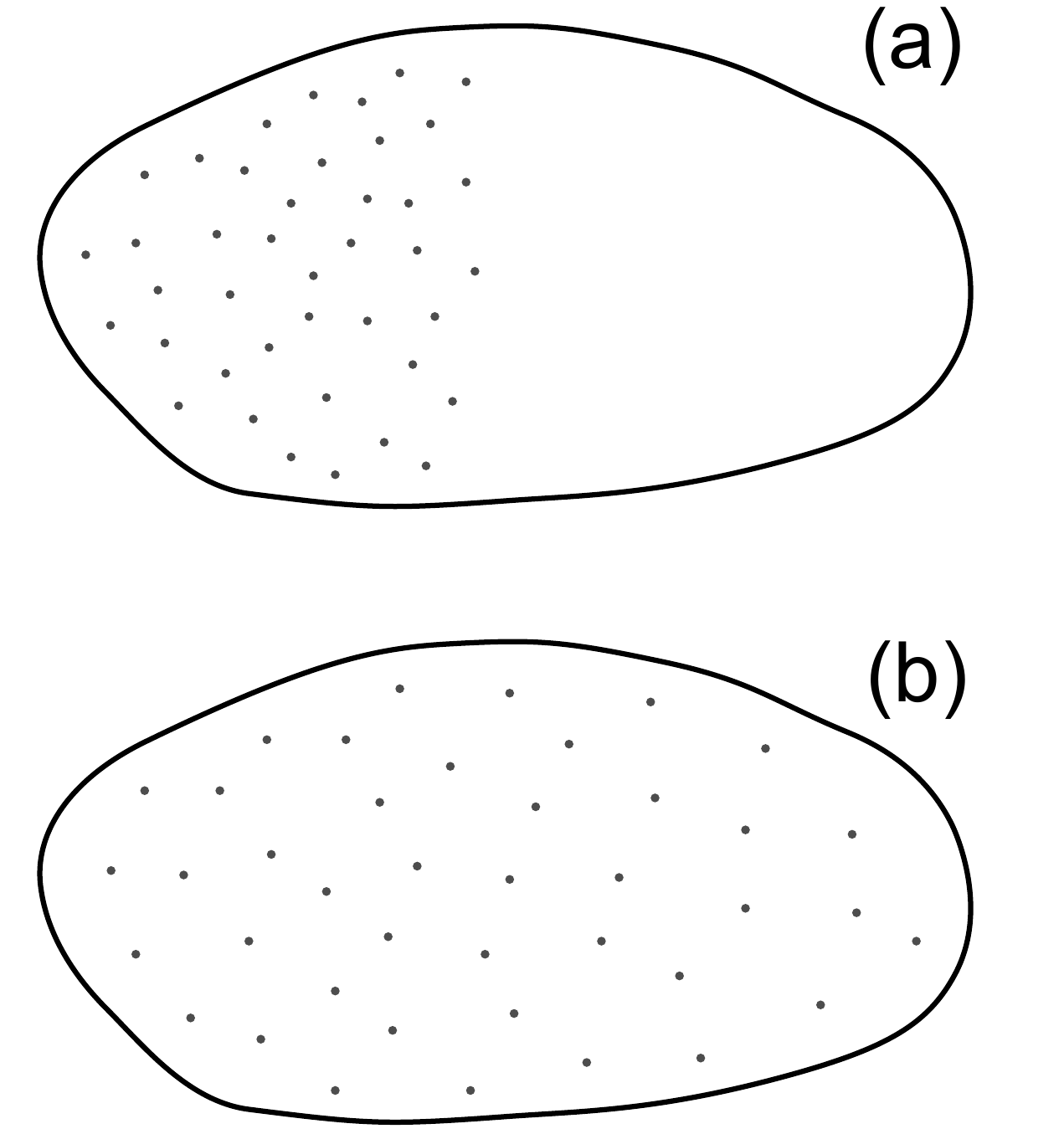}
\caption{Diffusion in a container with irregular boundary. The gas begins confined 
to the left half (a) and eventually spreads to fill the entire volume (b). 
This irreversible process occurs whether the gas is classical or quantum, 
and regardless of whether the particles interact.}
\label{fig:diffusion}
\end{figure}

Let us consider the gas diffusion illustrated in Fig.~\ref{fig:diffusion}, where an isolated container is initially half full.
To avoid complications of quantum indistinguishability, we assume that all the molecules in the gas have different masses.
The diffusion is clearly irreversible and can occur in the following four cases:
(1)~free classical gas;
(2)~classical gas with repulsive interaction;
(3)~free quantum gas;
(4)~quantum gas with repulsive interaction.
We focus on the quantum gas.
To characterize the irreversibility in the diffusion, we need an entropy that increases with the diffusion.

The vN entropy $S_{v}$  is evidently not such an entropy. Because the container is isolated, 
the entire diffusion is a unitary transformation. If the gas is initially in a pure quantum state, 
it remains pure forever, and $S_{v}$​ stays at zero---capturing nothing of 
the gradual diffusion or relaxation. The situation is essentially the same if the gas starts in a mixed state: 
although $S_{v}$​ is then nonzero, it still does not change, 
since unitary evolution preserves the spectrum of  $\hat{\rho}$. 
Consider the extreme case of noninteracting molecules that begin in a product state. Diffusion still proceeds, 
the single-particle wave functions spread to fill up the container\cite{Xiong_2011}, and the density becomes uniform. 
Yet the vN entropy of any single molecule, or of any collection of molecules, 
remains zero throughout, because no entanglement is generated. 
The irreversible process leaves no trace in $S_{v}$.

In contrast, the WvN entropy $S_{w}$ increases as the gas diffuses to fill the container, no matter whether the gas is interacting or not.
The reason is that the gas spreads in phase space during diffusion.
If there is no interaction, the spreading is in the spatial directions; if there is repulsive interaction, the spreading is in both momentum and position.
This means that $S_{w}$ can capture the irreversibility of the diffusion in Fig.~\ref{fig:diffusion}.

Importantly, the WvN entropy $S_{w}$ is extensive, whereas the vN entropy $S_{v}$ is not.
Consider a system consisting of two subsystems A and B.
One can show that in general, for the WvN entropy,
\begin{equation}
S_{w}^{\mathrm{A+B}} = S_{w}^{\mathrm{A}} + S_{w}^{\mathrm{B}}\,,
\end{equation}
when the dimensions of the Hilbert spaces of the system and the two subsystems are large enough~\cite{Han2015,GWvN}.
For the vN entropy $S_{v}$, similar results hold only in special cases.
And we know that thermodynamic entropy is extensive.

We have not noticed any results that contradict experiments when the vN entropy is used as thermodynamic entropy.
This is due to a mathematical coincidence for the case where $S_{v}$ is used most frequently.
In this case, a small system $\mathcal{S}$ interacts with a much larger system that is often referred to as a heat bath $\mathcal{B}$.
When the whole system $\mathcal{S}+\mathcal{B}$ is in a typical pure state, 
Page showed that $S_{v}^{\mathcal{S}}\approx\ln D_{\mathcal{S}}$, where $D_{\mathcal{S}}$ is the dimension of the Hilbert space of $\mathcal{S}$~\cite{Page}.
At the same time, one can also show that $S_{w}^{\mathcal{S}}\approx\ln D_{\mathcal{S}}$~\cite{GWvN}.
This means that $S_{w}^{\mathcal{S}} = S_{v}^{\mathcal{S}}$. In fact, 
in this case,  both entropies can be used to derive the Boltzmann--Gibbs distribution 
with the maximum-entropy principle~\cite{vonNeumannBook,GWvN}.

\section{Physical Basis}\label{sec:basis}

It has been discussed why the WvN entropy has been largely ignored~\cite{Goldstein2010}.
In our opinion, the biggest reason is that it was defined for a special set of orthonormal basis functions.
It was not clear why these basis functions are preferred or how to choose this kind of basis for general quantum systems, in particular for spin systems.
Let us first re-examine von Neumann's choice in 1929.

Von Neumann's choice is a set of wave functions $|w_{j}\rangle$ that are localized in Planck cells.
He used them to construct macroscopic position and momentum operators,
\begin{equation}
\hat{X} = \sum_{j} x_{j} |w_{j}\rangle\,, \qquad
\hat{P} = \sum_{j} p_{j} |w_{j}\rangle\,.
\end{equation}
Apparently, these two operators commute, $[\hat{X},\hat{P}]=0$.
Von Neumann argued that these two operators reflect the fact that ``in a macroscopic measurement 
of coordinate and momentum, really two physical quantities are measured simultaneously''~\cite{von2010proof}.
The cloud chamber, which is used to track the movement of electrons, positrons, and other 
microscopic particles, is precisely an instrument that measures a particle's position and momentum simultaneously.

We define a set of orthonormal basis functions as \textit{physical} if one can construct an instrument 
to make the corresponding measurement without using an exponentially large amount of physical resources.
A quantum computer is a good illustration of this definition.
For a given quantum computer with $n$ qubits, there is a set of computational basis states
\begin{equation}
|j\rangle = |j_{1} j_{2} \cdots j_{n}\rangle\,, \quad j=0,1,2,\dots,2^{n}-1\,,
\end{equation}
where $j_{1}j_{2}\cdots j_{n}$ is the binary representation of the integer $j$.
This basis is clearly physical, as has been demonstrated in many laboratories.
Now consider another basis related to $|j\rangle$ by a unitary transformation, $|u_{j}\rangle = \hat{U}|j\rangle$.
When the quantum computer is in a given quantum state and we would like to make a measurement 
with respect to $|u_{j}\rangle$, i.e., $|\psi\rangle = \sum_{j} a_{j} |u_{j}\rangle$, we 
need first to perform the unitary transformation $\hat{U}^{\dagger}|\psi\rangle$ and then make a measurement with respect to $|j\rangle$.
It is well known that one usually needs an exponentially large number of quantum gates 
to realize a typical unitary transformation~\cite{PRA2005,Sun2023}.
This shows that a physical basis is rare for a given quantum computer.

In the real world, the situation is similar.
There are infinitely many Hermitian operators; therefore, according to von Neumann~\cite{von2010proof}, 
there are as many measurements in principle.
In reality, we can make only a very limited number of measurements.
In particular, at the microscopic level, they include position, momentum, 
angular momentum, and spin.
For energy operators, we can make related measurements for systems 
of a small number of atoms.
When the system is big, e.g., with $10^{23}$ atoms, there is no general method to measure 
their energy levels. It is only possible for systems that have quasiparticle excitations.

For a given quantum system, if there are several different sets of physical basis functions, one can choose any one of them to 
compute the Wigner-von Neumann entropy.
It has been shown in Ref.~\cite{GWvN} that the results are almost the same for different choices.

The physical basis is intimately related to the preferred basis\cite{Prefer_rmp}. 
The subtle difference will be discussed elsewhere. 

\section{Conclusion}\label{sec:conclusion}

Von Neumann proposed two quantum entropies almost one hundred years ago to address the issue of how to define thermodynamic entropy for quantum systems.
From a modern perspective, both are important and useful.
However, the von Neumann entropy proposed in 1927 should be used as a measure of quantum entanglement, while the Wigner-von Neumann entropy proposed in 1929, which has been largely forgotten, should be used as the thermodynamic entropy for quantum systems.
Interestingly, using a specific set of basis functions to define quantum entropy was recently rediscovered in 2019~\cite{Aguirre2019}.

\begin{acknowledgments}
The author thanks Zhigang Hu and Xiaofeng Jin for helpful discussions. 
This work was supported by the National Natural Science Foundation of China (92365202, 12475011, 11921005), 
the National Key R\&D Program of China (2024YFA1409002), and the Shanghai Munici- pal Science and Technology Project (25LZ601100).
\end{acknowledgments}


\end{document}